\documentclass{appolb}
\usepackage{graphicx}

\begin{document}
\title{Pairing dynamics and time dependent density functional theory%
\thanks{Presented at the XXXV Mazurian Lakes Conference on Physics, Piaski, Poland, September
3--9, 2017}%
}
\author{P.~Magierski$^{a,b}$,  J.~Grineviciute$^{a}$, K.~Sekizawa$^{a,b}$
\address{$^{a}$Faculty of Physics, Warsaw University of Technology, ulica Koszykowa 75, 00-662 Warsaw, Poland \\ 
		$^{b}$Department of Physics, University of Washington, Seattle, WA 98195--1560, USA}}

\maketitle
\begin{abstract}
 We discuss issues related to pairing dynamics in nuclear large amplitude collective motion. The examples of effects which
 are not properly described within BCS theory are presented. In the second part we review properties of time-dependent density functional theory 
 (TDDFT) and in particular we discuss the time-dependent superfluid local density approximation (TDSLDA) starting from the stationary action principle.
\end{abstract}
  
\section{Remarks on pairing dynamics}

The theoretical description of atomic nuclei and nuclear systems in general requires superfluidity as a crucial ingredient. 
Although the size of the pairing gap in nuclear systems does not exceed $3$\% of Fermi energy, 
the influence of the pairing correlations on dynamics of medium 
or heavy nuclei is essential. As one of the best examples serves the nuclear induced fission process, which cannot be
understood without taking into account pairing correlations \cite{bulgac_2016}. 
Their role, both in the ground state as well as in excited states, has been studied and analysed
for decades (see, \textit{e.g.}, Refs.~\cite{ring,shimizu,bender,dean,hashimoto2013} and references therein).
Still, dynamical aspects of the pairing field in large amplitude nuclear motion are usually not taken into account.
To be precise, most of the effects related to superfluidity are described within the single-particle picture, where
only one aspect of the pairing field is manifested, namely, the appearance of the energy gap at the Fermi level.
Within the BCS theory it is interpreted as the energy associated with the Cooper pair formation. 
This clearly produces a noticeable effect for large amplitude collective motion, being responsible for decreasing 
the one-body dissipation. Indeed the collective energy dissipation can be traced back 
to the single-particle level crossings at the Fermi level.
If the pairing correlations are active, the crossings will disappear and consequently the probability of the 
single-particle excitation process will be decreased. The Landau-Zener formula tells us that it will decrease 
exponentially with the square of the pairing gap magnitude. Consequently the nuclear motion will become closer to the adiabatic
limit and sometimes it may even justify the usage of a single adiabatic potential energy surface which become effectively 
decoupled from other degrees of freedom.
This decoupling is also revealed in the behaviour of mass parameters which in the
vicinity of level crossings behave as $1/\Delta$.
It is therefore assumed that the pairing field is constant or changes adiabatically as a function of the density variations.
This simplified description of the pairing field in nuclear dynamics is quite common as it can be applied within a relatively simple theoretical framework. 
It reduces the manifestation of the pairing correlations to the single complex number---the pairing gap---which is determined 
through the nuclear density.
This is in general not correct, since dynamical aspects of the pairing field are then completely neglected.
It is believed, however, that the pairing field
dynamics will produce only small corrections to the commonly accepted
picture of nuclear dynamics. Moreover, the proper treatment
of the pairing field dynamics requires to use more advanced approaches
leading to a rapid increase of computational complexity. 

In order to understand better this distinction of the pairing treatments,
it is instructive to start with reminding the differences between two theoretical frameworks,
namely, BCS and Hartree-Fock-Bogoliubov (HFB) approaches.
Formally the difference originates from the fact that in the former case, the third transform
of the Bloch-Messiah decomposition (of the Bogoliubov transformation) is equal to unity \cite{bm}.
This requirement, which is responsible for a significant simplification,
has serious consequences in the pairing description. It means that one cannot describe processes due to 
the quasiparticle scattering and therefore the phenomena originating from the interaction of quasiparticles with a
nonuniform pairing field (scattering on the pairing potential).
These phenomena, although not so well pronounced in nuclear ground states, can have an impact on
the dynamics of nuclear systems. Hence, effects related to the nonuniformity
of the pairing field, such as the existence of Andreev states or Andreev reflection \cite{pm_2007, bulgac_2005}, 
which are well known in condensed matter physics, cannot be described within the so-called HF+BCS approach.
In this framework the pairing field is expressed as $\Delta({\bf r}) = g({\bf r})\sum_k v^{*}_{k} u_{k} |\psi_{k} ({\bf r})|^2$
and thus it resembles the density profile, even when $g$ is coordinate dependent.  
This is due to the fact that  the occupation numbers of HF orbitals are 
just numbers associated with each orbital, and therefore one cannot describe a configuration with
position dependent occupation numbers. It implies that  \textit{e.g.} a quantum vortex solution of HF+BCS
equations does not exist, as it requires variations of occupation numbers as a function of the distance
from the vortex core, where the system is normal \cite{yu_vortex}. This prevents applications of HF+BCS theory
to describe  \textit{e.g.} the inner crust of neutron stars, where various vortex-impurity configurations may exist
and their properties (pinning energies) are expected to be crucial for understanding the pulsar glitch phenomenon \cite{wlazlowski_2016}.

In the case of time dependent phenomena in nuclear systems the aforementioned limitations of the HF+BCS
approach lead to an effective ``freezing" of excited modes of the pairing field. Consequently
the nuclear dynamics is governed by the nucleon density evolution only, with the pairing gap
adjusting at each time to a given nuclear configuration. It is easily seen from the equations
\begin{equation} \label{hf}
i\hbar\frac{\partial}{\partial t}\psi_{k}({\bf r},t) = \hat{h} \psi_{k}({\bf r},t),
\end{equation}
which define the evolution of HF orbitals according to the mean-field $\hat{h}$.
They are used to set the basis for the evolution of the diagonal density matrix and
the pairing tensor (see Refs.~\cite{lacroix, ebata}):
\begin{eqnarray}
 \frac{d}{dt}\rho_{kk}(t) = \Delta_{k\bar{k}}(t)\nu_{k\bar{k}}^{*}(t)- \Delta_{k\bar{k}}^{*}(t)\nu_{k\bar{k}}(t), \\
 \frac{d}{dt}\nu_{k\bar{k}}(t) = \Delta_{k\bar{k}}(t)(1-2\rho_{kk}(t)) .
\end{eqnarray} 
Clearly the spatial dependence of the pairing field cannot be described within this framework.

It is instructive to consider the following  process:
suppose, we deal with a uniform system which is superfluid and the time evolution is triggered by an
external spatially modulated pairing field $\Delta_{\rm ext}({\bf r})$. 
Note that before the external field is switched on, the HF+BCS approach is equivalent
to HFB equations, since initially there is no quasiparticle scattering and the canonical basis
corresponds simply to plane waves.
However when the system is perturbed by the external pairing field the translational invariance
is lost and  the density waves may be excited. This process cannot
be described within TDHF+BCS treatment, as one can easily infer from Eq.~(\ref{hf}).
Namely, the system is initially described by $\psi_{k}({\bf r})\propto \exp(i {\bf k}\cdot {\bf r})$,
which are eigenstates of $\hat{h}$, and there is no mechanism to break the translational invariance
by the spatially modulated pairing field.
Thus the perturbation induced by the external pairing field will result in a modification of the
magnitude of the pairing gap only:
\begin{equation}
\Delta_{k\bar{k}} \rightarrow \Delta_{k\bar{k}} + \Delta^{\rm ext}_{k\bar{k}},
\end{equation}
leading to oscillations of the uniform pairing field.
Since the density reads:
\begin{equation}
\rho({\bf r}, t)=\sum_{k} \rho_{kk}(t)\,|\psi_{k}({\bf r},t)|^{2},
\end{equation}
the translational symmetry breaking may occur through the symmetry 
breaking terms in the mean-field Hamiltonian, but these are absent, according to our initial assumption.
Consequently the spatial modulation of the pairing field in the TDHF+BCS dynamics may be generated
only as a consequence of the evolution of the normal density $\rho$. Last but not least, it turns out that 
TDHF+BCS equations 
violate the continuity equation producing various unwanted effects (see Ref. \cite{lacroix}).

On the contrary the TDHFB framework offers a possibility to take into account excitation modes of the pairing
field $\Delta({\bf r},t)$ itself (\textit{e.g.} Bologliubov phonons).
These modes are treated on the same footing as the normal degrees of freedom described by 
$\rho({\bf r},t)$.
Within this approach an external, inhomogeneous pairing field will induce various processes in the initially
uniform system due to the quasiparticle scattering.
Two examples of results, where the pairing dynamics played a crucial role, comprise the induced fission of $^{240}$Pu \cite{bulgac_2016}
and the collisions of two superfluid nuclei \cite{magierski_2017,INPC,FUSION17}. In the former case the dynamics of the pairing field
causes much longer fission times than expected, based on the simplified pairing treatment. In the latter case the dynamics of the pairing field 
lead to the soliton-like excitation of the pairing field of two colliding nuclei resulting in the modification of the kinetic energy of the fragments 
and the capture cross section.

Summarizing, in this section we described differences between TDSLDA and TDHF+BCS-type approaches, discussing an example 
of the process which cannot be described within TDHF+BCS framework. Still the advantage of pure TDHF \cite{tdhf} or TDHF+BCS lies in
their relative simplicity and the description they offer is correct if magic nuclei or relatively high energies are considered.
There are indications however, that the induced fission or collisions of nonmagic nuclei may require to consider more advanced approach, which
takes pairing dynamics into account.

\section{Time-dependent density functional theory}

In this section we review briefly the developments in density functional theory (DFT) extended to superfluid systems, which
allow to overcome difficulties described in the previous section resulting from the incorrect treatment of pairing dynamics (see also Refs.~\cite{ARNPS__2013, Mag2016}).
DFT has become a standard theoretical tool as
it offers a universal and formally exact approach, which had enormous practical successes \cite{Kohn:1999,Dreizler:1990lr,Eschrig:1996,Parr:1989}. 
It is widely used in the field of condensed matter and in particular well suited to determine
properties of electronic systems \cite{Picket:1989,Brack:1993,Brivio:1999,Freysoldt:2014}. 

The case of atomic nuclei is more complicated, however, since two types of particles, neutrons and protons, need to be taken into account in the description of the system. Moreover the nuclear interaction involves many terms, 
including also the three-body force, without a clear recipe concerning its functional form. Consequently the nuclear energy density functionals have various forms, 
the most popular being the Skyrme functional, which despite of known shortcomings is
still widely used (see, \textit{e.g.}, Refs.~\cite{Tarpanov:2014,Rodriguez-Guzman:2014,Baldo:2013,Fayans:2001,unedf,ab2017} and references therein).
For nonsuperfluid systems the simple scheme offered by the energy density functional theory is very attractive, 
as instead of searching for the wave function of an 
$N$-particle system, which depends on $3N$ variables, one solves a system of $N$ nonlinear, coupled partial differential equations. 
It can be achieved through the application of the Kohn-Sham (K-S) scheme,
in which the interacting system is replaced by the equivalent (\textit{i.e.} of the same density distribution) noninteracting system
defined through the set of orbitals. These orbitals are in turn determined from variational principle \cite{KS65}, which
is equivalent to the minimization of the functional and generates the set of nonlinear equations defining the density distribution.
The formulation of DFT limits its applicability to the ground-state properties of the system.
In order to address the excited states and in particular nonequilibrium processes like nuclear fission or reactions 
an extension of the DFT is necessary to include the time evolution. 

This can be achieved through the time-dependent density functional theory (TDDFT), which can be used
to describe nonstationary situations in systems consisting of nuclei, atoms, molecules, solids, or nanostructures  
(see Refs.~\cite{Ullrich:2012,Marques:2012,Oni2002} and references therein). Whereas DFT is based on the Hohenberg-Kohn
theorem proving the existence of the unique density functional, TDDFT relies on the Runge-Gross mapping
which ensures that the evolution of the quantum system, \textit{i.e.} its wave function, can be determined through
the density (up to an arbitrary phase) \cite{Runge:1984mz}.
Despite these similarities with the static DFT, the time dependent theory may exhibit nonlocality in time, which leads to various problems related to causality principle. The so-called causality paradox has been resolved in a series of papers \cite{Rajagopal:1996,Leeuwen:1998,Leeuwen:2001,Mukamel:2005,Vig2008}.
Nonlocality in time is responsible for memory effects, which means that the behaviour of the system is dependent 
on the densities at earlier times \cite{Dob1997, Dob1997a}. This memory is, in principle, infinitely long-ranged
and very little is known about its behavior. This fact, together with a serious complication of resulting
time dependent equations, which would become integro-differential equations, results in the most common
approximation in TDDFT ignoring memory effects. The price which one pays for this simplification
is an incorrect treatment of energy dissipation processes \cite{Ullrich:2012}.

The existence of superfluidity and its incorporation in TDDFT leads to additional complications.
The first attempt to develop the formal framework of DFT for superconductors has been triggered
by the discovery of high-temperature superconductivity \cite{Oli1988, Wack1994}.  
Namely,  it can be achieved through the introduction of an anomalous density 
$\chi({\bf r}\sigma, {\bf r\rq{}}\sigma\rq{})=\langle \hat\psi_{\sigma\rq{}} ({\bf r}\rq{}) \hat\psi_{\sigma}({\bf r}) \rangle$
($\sigma$ denotes the spin degrees of freedom), which plays the role of the superconducting order parameter.
The pairing potential is then  defined as a functional  derivative of the energy functional with respect to $\chi$:
\begin{equation}
\Delta ({\bf r}\sigma, {\bf r\rq{}}\sigma\rq{})=\frac{\delta E(\rho, \chi)}{\delta \chi^{*}({\bf r}\sigma, {\bf r\rq{}}\sigma\rq{})}.
\end{equation}
Introducing Bogoliubov transformation (see below), which allows to express both normal and anomalous densities
in a form similar to the orbital expansion in conventional DFT,
one arrives at Kohn-Sham scheme for superfluid Fermi systems, which
formally resemble the Bogoliubov-de Gennes equations.
Unfortunately this set of equations is of the integro-differential form and therefore the above formulation
has rarely been used in practice.
This complication comes from the nonlocality of the pairing potential $\Delta ({\bf r}\sigma, {\bf r\rq{}}\sigma\rq{})$.
It turned out, however, to be possible to formulate the problem using a local pairing field \cite{Kur1999}. The justification for the so-called   
SLDA (Superfluid Local Density Approximation) has been developed in a series of papers 
(see Refs.~\cite{BY:2002fk,Bulgac:2002uq,BY:2003,Yu:2003,Bulgac:2007a, bulgac2013})
and was shown to be very accurate for nuclei and cold atomic gases. 
The prescription involves the renormalization of the pairing coupling constant, which is a function of the momentum cutoff.
In the case of the spherical cutoff the analytic formula can be derived (spin indices are omitted for clarity):
\begin{eqnarray}
\Delta({\bf r}) &=& -g_{\rm eff} ({\bf r})\chi_{c}({\bf r}), \\[2mm]
\frac{1}{g_{\rm eff} ({\bf r})} &=& \frac{1}{g ({\bf r})} - \frac{m k_{c}({\bf r})}{2\pi^{2}\hbar^{2}}
\left ( 1 - \frac{ k_{\rm F}({\bf r}) }{ 2k_{c}({\bf r}) } \ln \frac{ k_{c}({\bf r}) + k_{\rm F}({\bf r})  }{ k_{c}({\bf r}) - k_{\rm F}({\bf r}) }  \right )  ,
\end{eqnarray}
where anomalous density $\chi_{c}$ is defined within the truncated space and $k_{c}$ is the 
momentum cutoff.
This prescription works in the case of static DFT extended to superfluid system, but is only of a little help 
in the case of TDDFT, for the reasons which are discussed below (see also \cite{Mag2016,janina}). 

We may formulate TDDFT from the action stationarity principle, by defining the action (without including
memory effects):
\begin{equation} \label{action}
S=\int_{t_0}^{t_1}\left ( \langle 0(t)| i\frac{d}{dt}| 0(t) \rangle - 
    E\bigl[\rho({\bf r}\sigma,{\bf r'}\sigma', t),\chi({\bf r}\sigma,{\bf r'}\sigma', t)\bigr] \right) dt.
\end{equation}
The energy density functional in $E$ in principle also contains currents but they do not affect the derivations, so we will omit them for clarity.
In the above equation $|0(t)\rangle $ is a state which is a quasipartile vacuum $\alpha_{\mu}(t)|0(t)\rangle = 0$,
where $\alpha_{\mu}(t) = \hat{B}(t)a({\bf r}\sigma)\hat{B}^{\dagger}(t)$. 
It can be treated in the Kohn-Sham scheme (similarly as in the case of nonsuperfluid system), as a fictitious state describing an equivalent
non-interacting system having the same densities $\rho$ and $\chi$ as the true interacting system.
The operator $\hat{B}(t)$ defines the Bogoliubov transformation (and the state $|0(t)\rangle $) and
can be written as $\hat{B}(t)=\exp[i\hat G(t)]$, where $\hat{G}(t)$ is the hermitian operator of the form:
\begin{eqnarray}
& &\hat{G}(t)  =  
\int d{\bf r}d{\bf r}' \sum_{\sigma,\sigma'}h({\bf r}\sigma,{\bf r'}\sigma',t )a^{\dagger}({\bf r}\sigma)a({\bf r'}\sigma')  
- \frac{1}{2}\int d{\bf r} \sum_{\sigma}h({\bf r}\sigma,{\bf r}\sigma,t ) \\
              & + & \int d{\bf r}d{\bf r}' \sum_{\sigma,\sigma'} \left ( \frac{1}{2}\Delta^{*}({\bf r}'\sigma',{\bf r}\sigma,t )a({\bf r}\sigma)a({\bf r'}\sigma')  + \frac{1}{2}\Delta({\bf r}\sigma,{\bf r'}\sigma',t )a^{\dagger}({\bf r}\sigma)a^{\dagger}({\bf r'}\sigma') \right ). \nonumber
\end{eqnarray}
Although ${\bf r}$ is formally a continuous variable in practical applications one performs calculations on the lattice and thus it is discretized leading to sums instead of integrals
in the above formulas. Hence matrices $h$ and $\Delta$ define the matrix $G(t)$:
\begin{eqnarray} \label{tdslda}
G(t) = 
\left ( \begin{array}{cc}
h(t)&\Delta(t)\\
\Delta^{\dagger}(t)&-h^* (t)
\end{array} \right )  
\end{eqnarray}
which define the matrix of Bogoliubov transformation:
\begin{eqnarray} \label{bog}
B(t) = 
\left ( \begin{array}{cc}
U(t)& V^{*}(t)\\
V(t)& U^{*}(t)
\end{array} 
\right )  = \exp[iG(t)]
\end{eqnarray}
where amplitudes $U_{\mu}({\bf r}\sigma, t)$ and $V_{\mu}({\bf r}\sigma, t)$ ($\mu$\,th coloumn of $U$ and $V$, respectively) play the role of Kohn-Sham orbitals.
The matrix of the Bogoliubov transformation relates the new basis which define the state 
$|0(t)\rangle $ to the initial coordinate basis:
\begin{equation}
\vec{c} = B(t)\vec{\gamma}(t),
\end{equation}
where
\begin{eqnarray}
\vec\gamma(t) = \left  ( \begin{array} {c}
  \vec\alpha(t)\\  
  \vec\alpha^{\dagger}(t)\\ 
\end{array} \right ),\hspace{10mm}
\vec{c} = \left  ( \begin{array} {c}
  \vec{a} \\  
  \vec{a}^{\dagger} \\ 
\end{array} \right ).
\end{eqnarray}

Clearly, since $B^{\dagger}(t)B(t)=B(t)B^{\dagger}(t)=I$, variations of Eq.~(\ref{action}) with respect to $U$ and $V$ are not independent and the conditions:
\begin{equation} \label{stationarity}
\frac{\delta S}{\delta U_{\mu}({\bf r}\sigma,t)} = \frac{\delta S}{\delta V_{\mu}({\bf r}\sigma,t)} = 0
\end{equation}
are not going to produce the correct equation of motion which conserves the structure of the product state $|0(t)\rangle$, unless
certain constraints on variations are imposed. This can be achieved also by noticing that the functional (omitting spin indices for clarity) \\
$E(\rho,\chi) = E\bigl[ \sum_{\mu}V^{*}_{\mu}({\bf r}, t) V_{\mu}({\bf r'}, t),   
\sum_{\mu}V^{*}_{\mu}({\bf r}, t) U_{\mu}({\bf r'}, t) \bigr]$ has to be invariant
under the transformation:
\begin{eqnarray} \label{inv}
&\sum_{\mu}& V^{*}_{\mu}({\bf r}, t) V_{\mu}({\bf r'}, t) \rightarrow  \nonumber \\
& & \alpha \sum_{\mu} V^{*}_{\mu}({\bf r}, t) V_{\mu}({\bf r'}, t) + (1-\alpha) \sum_{\mu} (1 - U_{\mu}({\bf r}, t) U^{*}_{\mu}({\bf r'}, t)) \nonumber \\
&\sum_{\mu}& V^{*}_{\mu}({\bf r}, t) U_{\mu}({\bf r'}, t)  \rightarrow   \nonumber \\
& &\alpha \sum_{\mu} V^{*}_{\mu}({\bf r}, t) U_{\mu}({\bf r'}, t) - (1-\alpha) \sum_{\mu} U_{\mu}({\bf r}, t) V^{*}_{\mu}({\bf r'}, t)
\end{eqnarray}
for an arbitrary parameter $\alpha$. This invariance is a direct consequence of the completeness of the Bogoliubov transformation: $B(t)B^{\dagger}(t)=I$.
In order to get the proper equation of motion one needs to set $\alpha=1/2$ obtaining the symmetric form of the energy density functional. 
In the so-called local TDDFT, which is denoted as time dependent local density approximation (TDSLDA), one limits
to local expressions: $h({\bf r},{\bf r'},t ) \rightarrow h({\bf r},t ) $ and 
$\Delta({\bf r},{\bf r'},t ) \rightarrow \Delta({\bf r},t )$.
Consequently from the condition (\ref{stationarity}) one arrives at TDSLDA equations (omitting spin indices for clarity):
\begin{eqnarray} \label{tdslda}
i\hbar\frac{\partial}{\partial t} 
\left  ( \begin{array} {c}
  U_{\mu}({\bf r},t)\\  
  V_{\mu}({\bf r},t)\\ 
\end{array} \right ) =
\left ( \begin{array}{cc}
h({\bf r},t)&\Delta({\bf r},t)\\
\Delta^*({\bf r},t)&-h^* ({\bf r},t)
\end{array} \right )  
\left  ( \begin{array} {c}
  U_{\mu}({\bf r},t)\\
  V_{\mu}({\bf r},t)\\ 
\end{array} \right ),
\end{eqnarray}
where the relation between $E$ and $h$, $\Delta$ reads: 
$h({\bf r},{\bf r'},t) = \frac{\delta E}{\delta\rho ({\bf r'},{\bf r},t)}$ and
$\Delta({\bf r},{\bf r'},t) = \frac{\delta E}{\delta\chi^{*} ({\bf r},{\bf r'},t)}$.
In deriving the above formula the following property is used:
\begin{equation}
\langle 0(t)| i\frac{d}{dt}| 0(t) \rangle = 
\frac{1}{2}i\int d^{3} r \sum_{\mu} \left ( V_{\mu}({\bf r},t)\frac{\partial V_{\mu}^{*}({\bf r},t)}{\partial t}+U_{\mu}({\bf r},t)\frac{\partial U_\mu^{*}({\bf r},t)}{\partial t} \right ).
\end{equation}

Note, however, that these equations have been obtained from the stationary action principle under condition that the Bogoliubov transformation fulfills the completeness
relation. Otherwise one would not be able to define the new state $|0(t + \Delta t)\rangle$ from the previous one $|0(t)\rangle$.
However, when the energy cutoff is introduced, then only certain amplitudes $U_\mu$ and $V_\mu$ are taken into account. In such a case the expression for the energy density functional
$E$ ceases to be invariant under the transform (\ref{inv}) and the resulting equations are not correct.
It means that although formally one may still use Eq.~(\ref{tdslda}), evolving only selected amplitudes $U_\mu$ and $V_\mu$, it does not lead to a unique determination of the state $\big|0(t)\bigr>$.
One of the manifestation of this problem is the energy nonconservation which will occur during the evolution \cite{janina}.
The fact that the energy of the system is conserved is a trivial observation based on the form of the action (\ref{action}), but it is only the case when the Bogoliubov transformation
is properly defined according to Eq.~(\ref{bog}).
Therefore during the evolution on the spatial lattice all amplitudes of $U$ and $V$ need to be evolved, unless for short time evolutions (see Ref.~\cite{janina}).

\subsection*{Acknowledgments} 
We thank A.~Bulgac and G.~Wlaz{\l}owski for discussions and critical remarks.
Authors acknowledge support of Polish National Science Centre (NCN) Grants 
nos. DEC-2013/08/A/ST3/00708 and  UMO-2016/23/B/ ST2/01789. 
We acknowledge PRACE for awarding us access to resource Piz Daint based in Switzerland at Swiss National Supercomputing Centre (CSCS), decision No. 2016153479. 
We also acknowledge Interdisciplinary Centre for Mathematical and Computational Modelling (ICM) of Warsaw University for computing resources at Okeanos, grant No. GA67-14.


\begin{thebibliography}{99}
\bibitem{bulgac_2016} A. Bulgac, P. Magierski, K.J. Roche, and I. Stetcu, Phys. Rev. Lett. {\bf 116}, 122504 (2016). 
\bibitem{ring} P. Ring and P. Schuck, The Nuclear Many-Body Problem (Springer-Verlag, Berlin, 2000).
\bibitem{shimizu} Y.R. Shimizu, J.D. Garrett, R.A. Broglia, M.Gallardo, and E. Vigezzi, 
                  Rev. Mod. Phys. {\bf 61}, 131 (1989).
\bibitem{bender} M. Bender, P-H. Heenen, and P-G. Reinhard,
                 Rev. Mod. Phys. {\bf 75}, 121 (2003).
\bibitem{dean} D.J. Dean and M. Hjorth-Jensen, 
               Rev. Mod. Phys. {\bf 75}, 607 (2003).
\bibitem{hashimoto2013} Y. Hashimoto, Phys. Rev. C {\bf 88}, 034307 (2013).
\bibitem{bm} C. Bloch and A. Messiah, Nucl. Phys. {\bf 39}, 95 (1962).
\bibitem{pm_2007} P. Magierski, Phys. Rev. C {\bf 75}, 012803(R) (2007).
\bibitem{bulgac_2005} A. Bulgac, P. Magierski, and A. Wirzba, Europhys. Lett. {\bf 72}, 327 (2005).
\bibitem{yu_vortex} Y. Yu and A. Bulgac, Phys. Rev. Lett. {\bf 90}, 161101 (2003).
\bibitem{wlazlowski_2016} G. Wlaz{\l}owski, K. Sekizawa, P. Magierski, A.Bulgac, and M.M. Forbes, Phys. Rev. Lett. {\bf 117}, 232701 (2016)
\bibitem{lacroix} G. Scamps, D. Lacroix, G.F. Bertsch, and K. Washiyama, Phys. Rev. C {\bf 85} 034328 (2012)
\bibitem{ebata}  S. Ebata, T. Nakatsukasa, T. Inakura, K. Yoshida, Y. Hashimoto, and K. Yabana, 
                 Phys. Rev. C {\bf 82}, 034306 (2010).
\bibitem{magierski_2017} P. Magierski, K. Sekizawa, and G. Wlaz{\l}owski, Phys. Rev. Lett. {\bf 119}, 042501 (2017).

\bibitem{INPC} K. Sekizawa, P. Magierski, and G. Wlaz{\l}owski, PoS\textbf{(INPC2016)}214 (2017); arXiv:1702.00069.
\bibitem{FUSION17} K. Sekizawa, G. Wlaz{\l}owski, and P. Magierski, EPJ Web of Conf. {\bf 163}, 00051 (2017).
\bibitem{tdhf} A. S. Umar, M. R. Strayer, Comp. Phys. Comm. {\bf 63} 179 (1991); V. Blum, G. Lauritsch, J. A. Maruhn,
P.-G. Reinhard, J. Comp. Phys. {\bf 100} 364 (1992); F. Calvayrac, P.-G. Reinhard, E. Suraud, C. A. Ullrich,
Phys. Rep. {\bf 337} 493 (2000); C. Golabek,
C. Simenel, Phys. Rev. Lett. {\bf 103} 042701 (2009); C. Simenel, Eur. Phys. J. {\bf A 48}, 152 (2012); P. Goddard, P. Stevenson, and A, Rios, Phys. Rev. 
{\bf C 92}, 054610 (2015); Phys. Rev. {\bf C 93}, 014620 (2016).
\bibitem{ARNPS__2013} A. Bulgac, Ann. Rev. Nucl. Part. Sci. {\bf 63}, 97 (2013). 
\bibitem{Mag2016} P. Magierski,  in ``Progress of time-dependent nuclear reaction theory"
(ed. Yoritaka Iwata) in the ebook series: ``Frontiers in nuclear and particle physics"
(Bentham Science Publishers);  arXiv:1606.02225.
\bibitem{Kohn:1999} W. Kohn, {\em Nobel Lecture: Electronic structure of matter -- wave functions and density functionals}, Rev. Mod. Phys. {\bf 71}, 1253 (1999).
\bibitem {Dreizler:1990lr} R.M. Dreizler and E.K.U. Gross, {\em Density Functional
  Theory: An Approach to the Quantum Many--Body Problem}, Springer-Verlag, Berlin, (1990).
\bibitem{Eschrig:1996} H.  Eschrig, {\em The Fundamentals of Density Fucntional Theory}, B.G. Teubner Verlagsgesellschaft Stuttgart - Leipzig (1996).
\bibitem{Parr:1989} R.G. Parr and W. Yang, {\em Density-Functional Theory of Atoms and Molecules}, Oxford University Press, New York (1989).
\bibitem{Picket:1989} W.E. Picket, Rev. Mod. Phys. {\bf 61}, 433 (1989).
\bibitem{Brack:1993} M. Brack, Rev. Mod. Phys. {\bf 65}, 677 (1993).
\bibitem{Brivio:1999} G.B. Brivio and M.I. Trioni, Rev. Mod. Phys. {\bf 71}, 231 (1999).
\bibitem{Freysoldt:2014} C. Freysoldt, B. Grabowski, T. Hickel, J. Neugebauer, G. Kresse, A. Janotti, and C.G. Van de Walle, Rev. Mod. Phys. {\bf 86}, 253 (2014).
\bibitem{Tarpanov:2014} D. Tarpanov, J. Dobaczewski, J. Toivanen, and B.G. Carlsson, Phys. Rev. Lett. {\bf 113}, 252501 (2014).
\bibitem{Rodriguez-Guzman:2014} R. Rodriguez-Guzman and L.M. Robledo, Phys. Rev. C {\bf 89}, 054310 (2014).
\bibitem{Baldo:2013} M. Baldo, L.M. Robledo, P. Schuck, and X. Vi\,nas, Phys. Rev. C {\bf 87}, 064305 (2013).
\bibitem{Fayans:2001} S.A. Fayans and D. Zawischa, Int. J. Mod. Phys. B {\bf 15}, 1684 (2001).
\bibitem{unedf} S. Bogner {\em et al.}, arXiv:1304.3713.
\bibitem{ab2017} A. Bulgac, M.M. Forbes, S. Jin, R. Navarro Perez, and N. Schunck,  arXiv:1708.08771.
\bibitem{KS65} W. Kohn and L.J. Sham, Phys. Rev. {\bf 140}, A1133 (1965).
\bibitem{Ullrich:2012} C.A. Ullrich, {\em Time-dependent density-functional theory: concepts and applications}, Oxford University Press, (2012).
\bibitem{Marques:2012} {\em Fundamentals of time-dependent density functional theory}, Eds. M.A.L. Marques, N.T. Maitra, F.M.S. Nogueira, E.K.U. Gross, and 
                                           A. Rubio, Springer, Berlin (2012).
\bibitem{Oni2002} G. Onida, L. Reining, and A. Rubio, Rev. Mod. Phys. {\bf 74}, 601 (2002).
\bibitem {Runge:1984mz} E.~Runge and E.K.U. Gross, Phys. Rev. Lett. {\bf 52}, 997 (1984).
\bibitem{Rajagopal:1996} A.K. Rajagopal, Phys. Rev. A {\bf 54}, 3916 (1996).
\bibitem{Leeuwen:1998} R. van Leeuwen, Phys. Rev. Lett. {\bf 80}, 1280 (1998).
\bibitem{Leeuwen:2001} R. van Leeuwen, Int. J. Mod. Phys. A {\bf 15}, 1969 (2001).
\bibitem{Mukamel:2005} S. Mukamel, Phys. Rev. A {\bf 71}, 024503 (2005).
\bibitem{Vig2008} G. Vignale, Phys. Rev. A {\bf 77}, 062511 (2008).
\bibitem{Dob1997} J.F. Dobson, M.J. Brunner, and E.K.U. Gross, Phys. Rev. Lett. {\bf 79}, 1905 (1997).
\bibitem{Dob1997a}  G. Vignale, C.A. Ullrich, and S. Conti, Phys. Rev. Lett. {\bf 79}, 4878 (1997).
\bibitem{Oli1988} L. N. Oliveira, E. K. U. Gross, and W. Kohn, Phys. Rev. Lett. {\bf 60}, 2430 (1988). 
\bibitem{Wack1994} O.-J. Wacker, R. K\"ummel, and E.K.U. Gross, Phys. Rev. Lett. {\bf 73}, 2915 (1994).
\bibitem{Kur1999} S. Kurth, M. Marques, M. L\"uders, and E.K.U. Gross, Phys. Rev. Lett. {\bf 83}, 2628 (1999).
\bibitem {BY:2002fk} A.~Bulgac and Y.~Yu, Phys. Rev. Lett. {\bf 88}, 042504 (2002).
\bibitem {Bulgac:2002uq} A.~Bulgac, Phys. Rev. C {\bf 65}, 051305 (2002).
\bibitem {BY:2003} A.~Bulgac and Y.~Yu, Phys. Rev. Lett. {\bf 91}, 190404 (2003).
\bibitem {Yu:2003}  Y.~Yu and A.~Bulgac, Phys. Rev. Lett. {\bf 90}, 222501 (2003);
                    Phys. Rev. Lett. {\bf 90}, 161101 (2003)
\bibitem {Bulgac:2007a} A.~Bulgac, Phys. Rev. A {\bf 76}, 040502 (2007).
\bibitem{bulgac2013} A. Bulgac, Annu. Rev. Nucl. Part. Sci. {\bf 63}, 97 (2013).
\bibitem{janina} J. Grineviciute, P. Magierski, A. Bulgac, S. Jin, and I. Stetcu, In this issue of Acta Phys. Pol. B. 

\end{thebibliography}
\end{document}